\documentclass[twocolumn,showpacs,preprintnumbers,amsmath,amssymb,prl]{revtex4}
\usepackage{graphicx}% Include figure files
\usepackage{dcolumn}% Align table columns on decimal point
\usepackage{bm}% bold math
\usepackage{url}% url typeface

\newcommand\eq{\begin{equation}}
\newcommand\en{\end{equation}}
\newcommand\Da{{\rm Da}}

\newcommand\cH{{\cal H}}
\newcommand\cL{{\cal L}}
\newcommand\tripe{\tau_{{\text{ripen}}}}
\newcommand\tpore{\tau_{{\text{pore}}}}
\newcommand\trelax{\tau_{{\text{relax}}}}
\newcommand\mflat{\mu_{{\text{flat}}}}
\newcommand\csat{c_{{\text{sat}}}}
\newcommand\msur{\mu_{{\text{sur}}}}
\newcommand\msol{\mu_{{\text{sol}}}}
\newcommand\bx{\textbf{x}}
\newcommand\bn{\textbf{n}}
\newcommand\bk{\textbf{k}}
\newcommand\ceq{c_{\text{eq}}}

\begin{document}

\title{Ripening of Porous Media}

\author{Benny Davidovitch, Deniz Erta{\c s} and Thomas C. Halsey}  

\affiliation{Corporate Strategic Research, ExxonMobil Research and Engineering, 
1545 Route 22 East, Annandale, NJ 08801}

\date{October 22, 2002}

%%%%%%%%%%%%%%%%%%%%%%%%%%%%%%%%%%%%%%%%%%%%%%%%%%%%%%%%%%%%%%%%%%%%%%%%%%%
%%%%% Abstract
%%%%%%%%%%%%%%%%%%%%%%%%%%%%%%%%%%%%%%%%%%%%%%%%%%%%%%%%%%%%%%%%%%%%%%%%%%
\begin{abstract}
We address the surface tension-driven dynamics of porous media in nearly 
saturated pore-space solutions. We linearize this dynamics in the 
reaction-limited regime near its fixed points -- surfaces of constant 
mean curvature (CMC surfaces). We prove that the only stable interface for this dynamics 
is the plane, and estimate the time scale for a CMC surface to become unstable.
We also discuss the differences between open and closed system dynamics, pointing out the unlikelihood that CMC surfaces are ever realized in these systems on any time scale.              
\end{abstract}   

\maketitle
%%%%%%%%%%%%%%%%%%%
When solid grains are suspended in a solution saturated with the 
molecular constituents of the grains, they undergo coarsening   
under the thermodynamic driving force of surface tension.
During this phenomenon, known as Ostwald ripening, 
the free energy of the system is lowered by 
minimizing the contact area between the coexisting phases. Molecules 
dissolve from high-curvature areas of the 
interface, pass through the solution, and can precipitate in low-curvature 
surface regions. This dynamics can be very complicated and depends on many 
parameters 
such as chemical composition, induced temperature and pressure fields, 
and other 
factors.

In this Letter we study Ostwald ripening in porous media, where the fluid 
in the pore space is approximately saturated with the ingredients of the solid 
phase. An example would be a sedimentary material such as sandstone, with the 
water in the pore space saturated with the silica components of the 
rock. Another example would be crushed ice, with water vapor saturating the 
air in the pore space between ice grains. We identify the fixed points of 
dissolution-precipitation dynamics as surfaces of ``Constant Mean Curvature", 
(CMC surfaces), and then show quite generally that these surfaces are unstable 
fixed points of this dynamics.

If one suspends grains in a solution, a mean field theory due to 
Lifshitz-Slyozov and Wagner is successful in capturing the dynamics at 
late stages of the ripening process for low solid volume fraction \cite{lifshitz-slyozov}. 
Another example of surface tension driven ripening arises in the kinetics
of foams \cite{foams}. Unlike these examples, in typical porous media 
both the solid and pore space components of the medium are connected.

We thus pursue an approach more suited to porous media. 
Following \cite{Adler} we consider the evolution of an 
interface $\Gamma (t)\equiv\bx(u(t),v(t))$ between a porous, single 
component, isotropic solid and its ideal solution in the interstitial fluid. 
The solid is subject to a first 
order dissolution-precipitation reaction in a flow field of velocity ${\bf v}$. 
The normal velocity of the surface $u_n (\bx)$ into the pore space is given by
\begin{equation}
u_n (\bx) = -K_f \left( 1 - e^{-\frac{\Delta \mu (\bx) }{kT}} \right)~, 
\label{surfacedynamics1}
\end{equation}
where $K_f$ is the dissolution rate, and the precipitation rate is 
controlled by the Boltzmann factor associated with the difference 
$\Delta \mu (\bx) \equiv  \msur (\bx) - \msol (\bx)$
between the chemical potentials of solid and dissolved molecules at
the interface $\Gamma$. 
Referred to the chemical potential $\mflat$ for a flat surface in 
equilibrium with an ideal saturated solution of
concentration $\csat$, these are given by
\begin{eqnarray}
\msur (\bx) &=& \mflat + 2 \nu_m \sigma H(\bx) ~, 
\label{msurf} 
\\
\msol (\bx) &=& \mflat + kT \log \frac{c(\bx)}{\csat} ~,
\label{satchem}
\end{eqnarray} 
where $\nu_m$ is the molecular volume in the solid, $\sigma$ is the 
interfacial energy, $c({\bf x})$ is the concentration near the surface point $\bf{x}$, and $H(\bx)\equiv(1/2)(\delta S/\delta V)$ 
is the mean curvature, which measures the local variation in surface area 
$\delta S$ with respect to a volume change $\delta V$ of the solid. Here and elsewhere we define all concentrations with respect to the concentration in the solid.
For $\Delta \mu \ll kT$, Eq. (\ref{surfacedynamics1}) reduces to 
\begin{equation}
u_n (\bx) =
-K_f \frac{2\nu_m \sigma}{kT} \left [ H (\bx) - \frac{kT}{2\nu_m \sigma}
\log \frac{c(\bx,t)}{\csat} \right ].
\label{dissol}
\end{equation}
Since $H$ should be no larger than the inverse of a typical pore size $L$,
$H\sim L^{-1}$, and $\nu_m \sigma/kT$ is usually the inverse of a 
molecular scale,  
\begin{equation}
\frac{\nu_m \sigma}{kT} \gg H ~,
\label{scale}
\end{equation}
we see that the linearized form of the dynamics holds for a nearly saturated
solution $c\approx \csat$.
    
The transport of the concentration field $c$ in the 
solution is described by the advection-diffusion equation,
with the appropriate boundary condition at the solid interface \cite{Adler}. 
The velocity field satisfies the Navier-Stokes equation, with appropriate 
boundary conditions at infinity and at the solid surfaces inside the porous 
medium.

Since the full solution of these equations in a porous medium is daunting, 
we concentrate on two limiting cases of this problem, which we term 
``perfect open" and ``perfect closed" systems. In a perfect open system, the 
velocity of the flow through the medium is considered sufficiently 
rapid that in effect each element of the surface is in contact with fluid 
whose concentration $c$ of the solute is fixed by a distant reservoir. 
This criterion can be achieved by taking the limit of small Damk{\" o}hler 
number $\Da \equiv K_f/v \to 0$, with $v$ the average velocity of flow 
through the system \cite{daccord}. Most of this study will concentrate on this 
limit. The other limit of a ``perfect closed" system, in which $v=0$ and the 
diffusion constant $D \to \infty$, will be explored at the end of this Letter.

In the perfect open case, the concentration $c(\bx)$ on the solution side 
of the interface has the constant value $c_{\infty}$, and the 
interface dynamics is governed only by Eq. (\ref{dissol}), which
yields
\begin{equation}
u_n (\bx) = 
-K_f \frac{2 \nu_m \sigma}{kT}[H (\bx) - H_{\star}]~,
\label{surfacedynamics2}
\end{equation}
where $H_{\star} \equiv [kT/(2 \nu_m \sigma)] \log (c_{\infty}/\csat)$.
 
CMC surfaces, for which $H({\bf{x}}) = H_{\star}$ everywhere on the surface,
are the fixed points of the dynamics (\ref{surfacedynamics2}). 
These surfaces have been studied in a variety of contexts, notably for their 
relationship to certain phases of block co-polymers \cite{polymer}. 
The simplest examples are a sphere of radius 
$r=\frac{1}{H}$ and a cylinder of radius $r=\frac{1}{2 H}$.

A special case of CMC surfaces are ``minimal surfaces", for which $H(\bx) = 0$. 
These surfaces have been extensively studied in the context of analytic function 
theory, and many such surfaces have been discovered \cite{schwartzP}. For $H_{\star} = 0$,
Eq.~({\ref{surfacedynamics2}) is the equation of motion for a manifold 
seeking to minimize its own surface area, a dynamics that is sometimes called 
``motion under mean curvature" \cite{sethian}. 
Of particular interest in the context of porous media are the
triply periodic minimal surfaces, of which the Schwartz P-surface, shown in
Fig.~\ref{fig:schwartzp}(a), is a classic example. Note that both sides of the surface form
connected components, as in most porous media. Anderson \textit{et} al. 
have extended several such surfaces into CMC surfaces with $H \neq 0$ \cite{anderson}. 
The extension of the Schwartz P-surface to a family of CMC surfaces is shown 
in Fig.~{\ref{fig:cmc}, with an example depicted in Fig.~\ref{fig:schwartzp}(b). 
These surfaces vary continuously from the P-surface $(M)$ to a simple cubic 
lattice of barely touching spheres at one endpoint $(A)$, and to a simple cubic 
lattice of barely touching spherical holes 
at the other endpoint $(A')$ (see Fig.~{\ref{fig:cmc}}). 
Anderson \textit{et} al. also generated families corresponding to fcc, bcc, and diamond lattices of spheres, with qualitatively 
similar properties. Although we are aware of no reported
examples of ``amorphous" families of CMC surfaces, it is likely that such 
families exist \cite{preparation}. Such families would be most relevant 
to real porous media.

\begin{figure}
\includegraphics[scale=0.4]{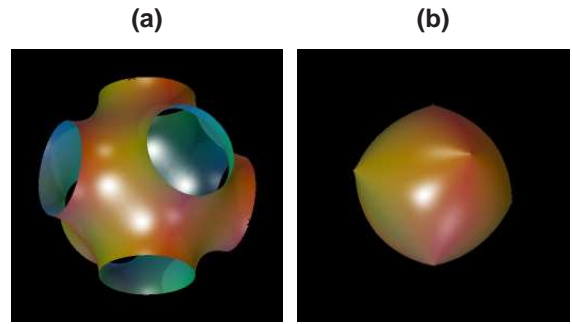}
\caption{\label{fig:schwartzp} (a) The Schwartz P-surface, a minimal (zero mean 
curvature) surface of simple cubic symmetry. A single unit cell is shown. 
(b) A single unit cell of size L of a CMC surface with $HL \approx 2.0$, which is close to a simple cubic lattice of 
touching spheres. Figures courtesy of J.T. Hoffman, MSRI Scientific 
Graphics Project, UC Berkeley \cite{msri}.}
\end{figure}

We now show that all CMC surfaces with the exception of planes
are unstable fixed points of Eq.(\ref{surfacedynamics2}). 
Consider a surface $\bx'(u,v) = \bx+\epsilon(\bx)\bn(\bx)$, whose deviation from
a CMC surface $\bx(u,v) \in \Gamma_\star$ with $H(\bx)=H_\star$ is given by a 
(small) normal displacement $\epsilon(\bf x)$.  
In \cite{preparation} we show that the corresponding variation 
$\delta H(\bx)=H(\bx')-H_\star$ is given to first order in $\epsilon$ by
\begin{eqnarray}
\delta H (\bx) &=& -\left[(2 H_\star^2 - K ) \epsilon + \nabla_s B \cdot 
\nabla_s \epsilon +
\frac{1}{2} {\nabla_s}^2  \epsilon \right], \label{perturbexpr}  \\ 
B(\bx) &\equiv& -\frac{1}{8} \log \left(H_{\star}^2 - K(\bx)\right),
\end{eqnarray}
where ${\nabla_s}$ and ${\nabla_s}^2$ are the surface gradient and Laplacian, 
respectively, and $K(\bx)$ is the Gaussian curvature at \bx. 
Substituting Eq. (\ref{perturbexpr}) in (\ref{surfacedynamics2}),
we obtain the linear dynamics near a CMC surface 
with mean curvature $H_{\star}$
\begin{eqnarray}
\frac{\partial \epsilon}{\partial t} &=& \frac{2 K_f \nu_m \sigma}{kT} \  
{\cal{L}}\epsilon, 
 \\
{\cal{L}} &\equiv& 
2 H_\star^2 - K +
\nabla_s B \cdot \nabla_s +
\frac{1}{2} {\nabla_s}^2 ~ .  
\label{linearspec}
\end{eqnarray}
Recall the definition of $H$ and $K$ in terms of the local principal radii of 
curvature of the surface $R_1$ and $R_2$,
\[
H = (R_1^{-1} + R_2^{-1})/2,
~;~~K = (R_1 R_2)^{-1}~.
\]
Since $2H^2 - K = (R_1^{-2} + R_2^{-2})/2 \geq 0$, 
uniform precipitation ($\epsilon >0$) 
decreases the curvature, making the surface locally more hospitable to deposition, 
while dissolution ($\epsilon < 0$) increases the curvature, favoring further 
dissolution. This observation already suggests that the surface is likely to be 
unstable.

\begin{figure}
\includegraphics[scale=.5]{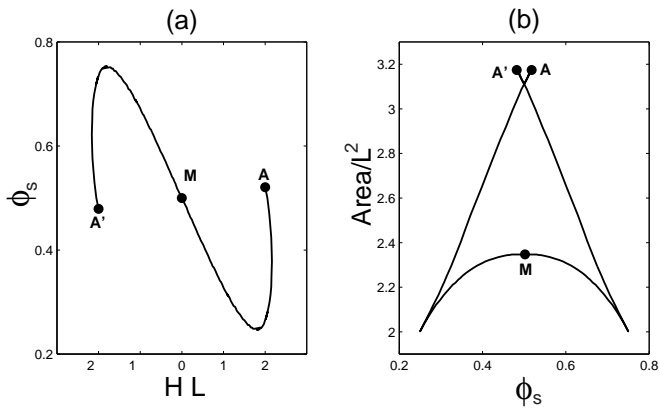}
\caption{\label{fig:cmc} (a) The curvature $HL$ versus volume fraction $\phi_s$ 
for the constant mean curvature extension of the Schwartz P-surface with 
unit cell size $L$. The 
minimal (H=0) P-surface is represented by the point $M$, and the endpoints 
correspond to (A) a simple 
cubic lattice of touching spheres, and (A') a simple cubic lattice of touching 
spherical holes. (b) The surface area $/L^2$ per unit cell of size $L$ versus enclosed solid volume fraction 
$\phi_s$ for the same simple-cubic CMC family. The lower branch is stable under 
periodic, volume-preserving, surface-minimizing dynamics. Figure adapted from 
Ref. \cite{anderson}, courtesy of Advances in Chemical Physics.}
\end{figure}

Because the term in $\cL$ which couples Gaussian curvature gradients with the gradient of $\epsilon$ is 
non-Hermitian, it is more convenient to work with the Hermitian
operator ${\cal H}$ defined via the ``gauge transformation"
\eq
\cH \equiv e^{B(\bx)} {\cal L} e^{-B(\bx)}
\en
so that
\begin{eqnarray}
\cH &=& V(\bx) + \frac{1}{2} {\nabla_s}^2 \ ,
 \\
V(\bx) &=& 
2H_{\star}^2 - K - \frac{1}{2}[\nabla_s^2 B + (\nabla_s B)^2]\ .  
\label{reducedform}
\end{eqnarray}
The eigenvalues of ${\cal H}$ are identical to those of 
$\cL$ \cite{non-Hermitian}, and they are bounded from above 
provided that $V(\bx)<\infty$. 

For any function $\Psi(\bx)$ defined on the surface, a lower bound on 
the maximum eigenvalue $\omega_0$ of $\cal H$ (consequently $\cal L$) 
can be established using the inequality
\begin{equation}
\omega_0 \ge \frac{\int_\Gamma dS \overline \Psi ({\cal H} \Psi)}
{\int_\Gamma dS \left|\Psi\right|^2} ,
\label{varbound}
\end{equation}
where $\int_\Gamma dS $ is the ordinary surface integral.
Using the trial function $\Psi_B(\bx) = e^{B(\bx)}$, we obtain
\begin{eqnarray}
\omega_0 &\ge& \frac{\int_\Gamma dS (H_{\star}^2 - K)^{-1/4}(2 H_{\star}^2 - K)}
{\int_\Gamma dS (H_{\star}^2 - K)^{-1/4}} \nonumber \\ 
&=& H_{\star}^{2} + \frac{\int_\Gamma dS (H_{\star}^2 - K)^{3/4}}
{\int_\Gamma dS (H_{\star}^2 - K)^{-1/4}}~.
\label{inner}
\end{eqnarray}
However, since $H^2 - K = [(1/R_1)-(1/R_2)]^2/4$ is non-negative everywhere 
on the surface and reaches zero homogeneously on $\Gamma$ only for a planar 
surface, we conclude that {\emph {the only marginally stable fixed point of the 
dynamics (\ref{surfacedynamics2}) is the plane}}. There are no other surfaces of 
any kind, porous or non-porous, that are metastable under Ostwald ripening. 
This conclusion applies both to periodic and to possible amorphous CMC surfaces. 
For a system with a pore scale of $L$, we expect this instability to manifest 
itself on a time scale of
\begin{equation}
\tripe = \frac{kT L^2}{K_f \sigma \nu_m}~.
\label{ripen}
\end{equation}
Note that intuitively, the instability is enhanced by an increase in the magnitude 
either of the mean curvature, or of the Gaussian curvature (since the latter is 
on average negative for a surface of large genus, such as a typical porous surface.)

For triply periodic CMC surfaces, the periodicity of the ``potential'' 
$V(\bx)$ implies that the eigenmodes $\Psi_{\bk,n}(\bx)$ of the 
operator $\cH$ (and of $\cL$) are surface Bloch functions,
\begin{eqnarray}
\cH \Psi_{\bk,n}(\bx) &=& 
\omega_{\bk,n} \ e^{i {\bk\cdot\bx}} U_{\bk,n} 
(\bx)
\label{spectrum} \\
 U_{\bk,n} (\bx) &=& U_{\bk,n} (\bx + \textbf{R})  
\label{defineBloch}
\end{eqnarray}   
where $\bf{R}$ is a lattice vector, ${\bf k}$ is the crystal momentum 
(confined to the first Brillouin zone), and $n$ is a discrete label 
distinguishing between different branches.
Standard techniques should allow determination of the spectrum 
of $\cH$ throughout the first Brillouin zone \cite{preparation}. 

As a concrete example, we have computed the positive 
eigenvalue $\omega_{0}$ corresponding to the $\bk = \textbf{0}$ 
eigenmode $\Psi_0 (\bx) \equiv U_{{\bf 0},0} (\bx)$ for the 
Schwartz P-surface numerically by maximizing the functional on the
RHS of Eq.~(\ref{varbound}). The details of the maximization procedure 
will be presented elsewhere \cite{preparation}. We obtain 
$\omega_0 \approx 6.1$ for a unit cell of size 
$L=1$, whereas the lower bound implied by Eq.~(\ref{inner}) is $\omega_0 \ge 2.9$. 

Since the eigenfunction of the largest eigenvalue has the same sign
(precipitation or dissolution) everywhere on the surface,
the unstable dynamics of Eq.~(\ref{surfacedynamics2}) involves transport 
of solute into (or out of) the pore space from the distant reservoir. 
Therefore, the nature of the dynamics might be different in a closed system, 
where the total amount of the solid component is conserved. 

In a ``perfect closed" system, we further regard the surface motion as 
reaction limited, such that the solute concentrations near the 
entire surface are approximately equal to the average concentration in 
the pore $\bar c$. This condition is satisfied at the pore scale 
$L$ if $D \gg K_f L/\csat$. By integrating Eq.~(\ref{dissol}), we see that $\bar c$ must then approach a value $\ceq$ determined by 
\begin{equation}
\log{\frac{\ceq(\bar H)}{\csat}} = \frac{2 \nu_m \sigma \bar H}{kT}~,
\label{Hconstrain}
\end{equation}
controlled by the average mean curvature $\bar H$ of the surface, over
a characteristic time scale $\tpore = L \csat/K_f$.  
For $\tau > \tpore$, the solute 
concentration $\bar c$ is ``slaved" to $\ceq(\bar H)$. Conservation of 
the total solid requires
\eq
\phi_t = \phi_s + {\bar c}(1-\phi_s)~, \label{phiconstrain}
\en
where $\phi_s$ is the solid volume fraction enclosed by the surface and
$\phi_t$ would be the solid volume fraction if none of the solid were 
dissolved. 
Eliminating ${\bar c} =\ceq(\bar H)$ 
from Eq.~(\ref{phiconstrain}) using Eq.~(\ref{Hconstrain}) 
shows that $\phi_s$ is very insensitive to the value
of $\bar H$,
\eq
\left|\frac{d \log \phi_s}{d \log \bar H}\right|\approx\frac{2\csat(1-\phi_t)}
{\phi_s(1-\csat)^2}\frac{\nu_m \sigma \left|\bar H \right|}{kT} \ll 1~,
\en
where we have used Eq.~(\ref{scale}). In a ``perfect closed" system, we take $\phi_s$ constant, which is an excellent approximation for $\tau > \tpore$.  This dynamics is now given by Eq.~(\ref{surfacedynamics2}) 
with the constant $H_\star$ replaced by $\bar H$, whose evolution with time is controlled by the constraint of constant $\phi_s \approx (\phi_t-\csat)/(1-\csat)$. This is surface area 
minimization under the constraint that the volume 
contained within the surface is conserved. 
 
Consider again periodic surfaces of the Anderson type. Figure \ref{fig:cmc}(b) 
shows the total surface area as a function 
of $\phi_s$ along the manifold of CMC surfaces of simple cubic structure. 
Although all CMC solutions represent fixed points of the 
dynamics, we anticipate that for a given $\phi_s$, only the CMC surface with 
the lowest area can be stable under these dynamics, and only if the 
system is also required to maintain its spatial periodicity. We have confirmed this
by direct simulation using Surface Evolver\cite{evolver,preparation}. This result is an extension of the 
well-known stability of the minimal surface (point $M$ in 
Fig.~\ref{fig:cmc}) to the entire
lower branch in Fig.~{\ref{fig:cmc}(b),
under periodic surface area minimization dynamics with conserved volume. 
A corollary of this result is that along this lower branch, 
$\cL$ has only one unstable $\bk=\textbf{0}$ mode \cite{preparation}, 
which is disallowed by solid conservation under closed system dynamics. Thus the stable mode with the longest relaxation time $\trelax$ will control approach to the stable CMC surface; 
dimensional analysis suggests that $\trelax \sim \tripe$.  

Nevertheless, for an extended system this apparent stability is compromised 
by the unstable hydrodynamic $(\bk \neq 0)$ modes in the eigenvalue spectrum 
Eq.~(\ref{spectrum}), which do conserve the overall $\phi_s$. 
It is possible for these unstable modes to be restricted 
to the topmost ($n=0$) band and to wavevectors $k < \pi/ \ell_0$,
such that transport of solute over length scales $> \ell_0$ is needed
to activate these modes. For amorphous surfaces, the band structure picture is not appropriate, but we still expect unstable modes to appear above a characteristic length scale $\ell_0 \gtrsim L$. In either case, it is natural to ask if there is a window 
of time over which the CMC surfaces might be observable, before the 
instability manifests itself.

During the time $\trelax$ characterizing the relaxation to the CMC surface, the diffusion length over which solid transport is possible is $\ell_D = \sqrt{D \trelax}$. Using the reaction-limited constraint $D \gg K_f L/\csat$ and $\trelax \sim \tripe$, with $\tripe$ given by Eq.~(\ref{ripen}), we find
\eq
\frac{\ell_D}{L} \gg \sqrt{\frac{kT L}{\sigma \nu_m \csat}} \gg 1 
\en
where the last inequality follows from Eq.~(\ref{scale}). Since we thus expect $\ell_D \gg \ell_0$, by the time a surface relaxes on the pore scale to a CMC surface, the diffusive transport between pores will have activated the unstable dynamics, taking the system away from this surface. Thus we do not expect an intermediate-time window over which CMC surfaces would be seen \cite{foot}.

Using representative values for $K_f$ and $\sigma$ for quartz in 
water \cite{quartz-rates}, we get for $L \sim 100$ $\mu$m that
$\tripe \sim $~500 million years, a time scale over which geological systems should be regarded as open. However, a similar estimate for 
limestone yields a value of $\tripe \sim $~7 years, which is quite short on geological time scales, so that these systems will correspond more closely to closed systems \cite{lime-rates}. Of course for real geological 
systems these processes involve complicated 
multicomponent equilibria, and other mechanisms such as pressure solution 
may dominate as well \cite{pressure-solution}. Nevertheless, there are 
probably geological circumstances under which surface-tension 
driven phenomena will be important.

We would like to thank I. Androulakis, P. Chaikin, S. Milner, and T. Witten for 
helpful and informative discussions. A. Herhold and R. Polizzotti advised us 
on the properties of real sedimentary materials. K. Brakke assisted us with 
questions regarding the use of the Surface Evolver software. We are grateful 
to F. Leyvraz for pointing out an error in our reasoning, and to M. Hastings 
for bringing to our attention the ``gauge" transformation used in 
Ref.~\cite{non-Hermitian}.

%\bibliography{porous}

\begin{thebibliography}{19}
\bibitem{foot}
\bibinfo{note}{This is only the case for connected pore spaces. Disconnected pore spaces, e.g., isolated spherical ``vugs", can obviously be stable.}
\end{thebibliography}

\end{document}